\documentclass[draft,jgrga]{agu2001}
\usepackage{graphicx}
\authorrunninghead{JI, KULSRUD, FOX, AND YAMADA}

\titlerunninghead{OBLIQUE ELECTROMAGNETIC DRIFT INSTABILITY}

\authoraddr{Hantao Ji, Russell Kulsrud, and Masaaki Yamada,
Center for Magnetic Self-organization in Laboratory and Astrophysical Plasmas,
Plasma Physics Laboratory, Princeton University, Princeton, New Jersey 08543, USA.
(hji@pppl.gov, rmk@pppl.gov, myamada@pppl.gov)}

\begin{document}

%
%
%
%
%

\title{An Obliquely Propagating Electromagnetic Drift Instability in the Lower Hybrid Frequency Range}

\author{Hantao Ji, Russell Kulsrud, William Fox, Masaaki Yamada}
\affil{Center for Magnetic Self-organization in Laboratory and Astrophysical Plasmas,
Plasma Physics Laboratory, Princeton University, Princeton, New Jersey, USA}

\begin{abstract}
By employing a local two-fluid theory, we investigate an obliquely propagating
electromagnetic instability in the lower hybrid frequency range driven by cross-field current or
relative drifts between electrons and ions. The theory self-consistently
takes into account local cross-field current and accompanying pressure gradients.
It is found that the instability is caused by reactive coupling between the backward
propagating whistler (fast) waves in the moving electron frame and the forward propagating 
sound (slow) waves in the ion frame when the relative drifts are large. 
The unstable waves we consider propagate obliquely to the unperturbed 
magnetic field and have mixed polarization with significant electromagnetic components. 
A physical picture of the instability emerges in the limit of large wavenumber characteristic
of the local approximation. The primary positive feedback mechanism is based on 
reinforcement of initial electron density perturbations by compression of 
electron fluid via induced Lorentz force. The resultant waves are qualitatively
consistent with the measured electromagnetic fluctuations in reconnecting
current sheet in a laboratory plasma.
\end{abstract}

\begin{article}

\section{Introduction}
Current-driven instabilities with frequencies higher than ion cyclotron frequency
($\omega > \Omega_i$) or wavelengths shorter than ion skin depth ($k\lambda_i >
1; \lambda_i \equiv c/\omega_{pi}$) have been a
 popular subject for space and laboratory plasma
research~\citep[see {\it e.g.}][]{gary93}. Recently, this topic has been revisited
in the context of magnetic reconnection~\citep[see {\it e.g.}][]{biskampbook},
where intense current density exists locally in the diffusion region.
In particular, the Lower Hybrid Drift Instability~\citep{krall71} (LHDI)
 driven by a density
gradient has received considerable attention as a potential source of anomalous
resistivity.

When the LHDI propagates nearly perpendicular to the magnetic field it is purely
electrotatic.
Such waves   have been observed at the low-$\beta$ edge of the current sheet
in the laboratory~\citep{carter02a}, in  numerical
simulations~\citep[see {\it e.g.}][]{scholer03},
and in space~\citep{shinohara98,bale02}.
They are driven unstable by inverse Landau damping of the drifting electrons.

However,
these electrostatic modes are largely stabilized~\citep{davidson77}
inside the high-$\beta$ reconnection layer, where the magnetic field gradient is large
and the $\nabla B $ drift of the electrons is in the wrong direction
to amplify the waves.  Further,
it is observed that their  amplitudes do not correlate with the fast reconnection
in the Magnetic Reconnection Experiment or MRX~\citep{carter02b}.
By  contrast, magnetic fluctuations up to the lower
 hybrid frequency range have been more recently detected~\citep{ji04a} in  this
 high-$\beta$ center of the current sheet in the MRX. These propagate
obliquely to the magnetic field, and their amplitudes exhibit positive correlations
with fast reconnection.
A theoretical explanation for  the origin of these magnetic fluctuations, 
other than the electrostatic perpendicularly propagating LHDI waves, 
is therefore in order.

Earlier, motivated by observations of high frequency magnetic fluctuations
 in a magnetic shock experiment,
Ross attempted~\citep{ross70} the first theoretical exploration of
 such candidate
obliquely propagating electromagnetic
high-frequency waves driven by a relative drift between electrons and
ions associated with
local currents.
Based on a two-fluid formalism in the electron frame, Ross showed that
unstable waves propagating obliquely to the magnetic field are
excited by reactive coupling between ion beam and whistler waves.
Such an instability is generally known as the
Modified Two Stream Instability~\citep{mcbride72,seiler76} (MTSI)
since it is driven by a local current across a magnetic field
unrelated to a diagmagnetic drift.

Extensions to a full kinetic treatment of both ions and electrons 
were made for this instability~\citep{lemons77,wu83,tsai84}.
Unlike the perpendicular LHDI, the obliquely propagating
MTSI persists in high-$\beta$ plasmas, where
the critical values of relative drift for the instability are typically a few
times the local Alfv\'en velocity, and possesses significant electromagnetic
components. However,
in most of these works, a finite pressure gradient self-consistent with the
cross-field current was left out in the wave dynamics. This neglect throws
doubt on the applicability of the MTSI to the MRX, where all the current
is due to inhomogeneities.

Recently, global eigenmode analyses~\citep{daughton99,yoon02,daughton03} of
the current driven instabilities have been carried out to take into
account the effects of boundary conditions of a Harris current
sheet~\citep{harris62}.
This followed earlier work on the same subject~\citep{huba80}.
It was found that for short wavelengths ($k\lambda_e \sim
1; \lambda_e \equiv c/\omega_{pe}$), the unstable modes concentrate
at the low-$\beta$ edge, and they are predominantly electrostatic similar to
the perpendicular propagating LHDI.
In contrast, for relatively longer wavelengths ($k\sqrt{\lambda_e
\lambda_i} \sim 1$), unstable modes with significant electromagnetic components 
develop in the center region.
These are similar to the MTSI at high-$\beta$.
For even longer wavelengths ($k\lambda_i \sim 1$), a drift kink
instability~\citep{daughton99}
is known to exist but this has a slower growth rate at more realistic ion-electron mass
ratios.
More recently, these analyses have been further extended to non-Harris
current sheets~\citep{yoon04,sitnov04}. When relative drift between electrons
and ions is enhanced, the central region is clearly dominated by instabilities
resembling the MTSI.

The first numerical simulations of the MTSI have been carried out in a two-dimensional local
model~\citep{winske85}, but focused on the electron heating.
Particle simulations have also been carried out in three-dimensions 
to study stability of a Harris current sheet under various but limited
conditions~\citep{horiuchi99,lapenta02,daughton03,scholer03,shinohara04,ricci04}.
It was found that at first the LHDI like instabilities are active only
at the low-$\beta$ edge, and modify the current profile
which then leads to the long wavelength electromagnetic modes, such as
drift kink instabilities or Kelvin-Helmholtz instabilities~\citep{lapenta02}.
Recent simulations using more realistic parameters (larger mass ratios with more
particles) in larger dimensions indicate \citep{ricci04} that
the MTSI-like modes also develop in the central region.
While the characteristics of the observed waves in the MRX current sheet
are generally consistent  with these linear stability analyses and nonlinear
simulation results, there has been yet no convincing physical explanation of
the observed electromagnetic waves in the lower hybrid frequency range.
Comparisons between MTSI and LHDI, the latter of which involves a self-consistent pressure gradient, 
were made based on local kinetic theories~\citep{hsia79,yoon94,silveira02}, but with a focus
on nearly perpendicularly propagating waves.
Extensions to larger propagation angles were also attempted 
earlier~\citep{zhou83,zhou91} but with few discussions on the underlying physics.

Motivated by the observations in the MRX and these recent theoretical
developments, we investigate this instability
based on a local two-fluid formalism in this paper. Our analysis is of the
MRX and includes the self-consistent pressure gradient with large propagation angles.
A local treatment is justified if the wavelength is short ($k\lambda_i \gg 1$)
and the growth rate is large ($\gamma \gg \Omega_i$),
compared to the global eigenmode analyses extending throughout the current layer 
(see for example \cite{kulsrud67}).
Our focus here is to reveal the underlying physics of the instability by
using the simplest possible model rather than to carry out more involved calculations.
We find that when the relative drifts are large, the instability is caused
by a reactive coupling between the backward
propagating whistler (fast) waves in the moving electron frame and the
 forward propagating
sound (slow) waves in the ion frame.
The unstable waves have a mixed electromagnetic character with both electrostatic and
magnetic components. They propagate obliquely to the unperturbed
magnetic field.
The primary positive feedback mechanism for the instability is identified as
reinforcement of initial electron density perturbations by an induced Lorentz force.
The role this instability plays in magnetic reconnection, such as
anomalous resistivity and heating, will be discussed in
a forthcoming paper~\citep{kulsrud04} that is based on quasi-linear
theory (see also \cite{winske85,basu92,yoon93}.)

\section{Theoretical Model}

The basic features of our model are described in this section. Since our main objective
is to understand physics of the underlying instability, we develop a theoretical
model, which contains the essential ingredients for the instability, yet remains simple enough so that
the feedback mechanism can be understood. In contrast to the past work, most of which 
is based on full kinetic theory, we find that we are able to use a simple two-fluid theory
and still obtain reliable results.  We show that
most features of the instability can be revealed by this simple model.

\subsection{Method of the calculation}

We wish to treat the  LHDI mode by an approach
somewhat different from earlier approaches.  Our basic
assumption is that the drift velocity is produced by equilibrium
gradients (LHDI) rather than an ion beam (MTSI).
In the MRX the gradients are the origin of the relative drift velocity
of the ions and electrons which is just the diamagnetic currents,
so that the instability is an LHDI.  However, since the
LHDI has been usually treated as a nearly perpendicular propagating
mode and we restrict ourselves in this paper to propagation at angles
finitely different from $ 90 $ degrees, we refer to our instability as
the obliquely propagating LHDI or more briefly the oblique LHDI.

	The reason we do not consider the LHDI near 90 degrees
is that it has been shown to be stable in the central
regions of the MRX and, as discussed in the introduction,
we are interested in explaining the observed instabilities
there.  In fact, as shown by \cite{carter02b}, the
gradient of the magnetic field is large there and the $\nabla B$
drifts cause the resonant electrons to drift in the opposite
direction than inferred from their current.  The oblique instability
we investigate is a non resonant one.

	We assume that the mode is at a large frequency compared
to the ion cyclotron frequency, and the wave length is
small compared to the ion gyration radius, so that the ions may
 be considered to be unmagnetized.  We also assume that the frequency
is small compared to the electron cyclotron frequency, $ \Omega_e $,  and that the
wave length is large compared to the electron gyration radius,
$ \rho_e $,  so that the
electron can be treated by the drift kinetic theory.  This theory is
described in Appendix A1, but the upshot of it is
that one expands  the Vlasov equation in the small parameter
$ \rho_e/ \lambda $ where $ \lambda $ is the perpendicular scale of the perturbation
as well as the equilibrium. One solves the Vlasov equation
to lowest order to obtain the zero order electron distribution
function $ f_0 $  from which one can obtain the electron pressure
tensor, $ {\bf  P}_e $.  Then one calculates the perpendicular
velocity moment of the first order distribution function $ f_1 $,
to find the perpendicular electron current.  But this calculation
is equivalent to taking the perpendicular electron fluid equation of
motion with this pressure tensor.
The parallel current is then obtained from the continuity
equation, $ \nabla \cdot {\bf  j}^e  - \partial (n_{e1}e) /\partial t = 0 $.

This procedure is totally equivalent to previous  calculations
giving identical results in the small $ \rho_e $ limit.  It might
be argued that one should consider waves with $ k_{\perp} \rho_e \sim 1$ 
since in previous work on the perpendicular LHDI the maximum
growth occurs when $ k_{\perp} \rho_e \sim 1 $.  However,
for the oblique LHDI the maximum growth actually occurs when
$ k_{\perp} \rho_e \ll 1$ and the mode becomes stable
for $ k_{\perp} \rho_e $ that approaches unity. (The guiding center treatment is appropriate for
inhomogeneous systems, since it makes no assumption about {\it near}
homogeneity and avoids the complicating approximations concerning it that
are usually made.)

It turns out that it is not appropriate to treat the
pressure tensor as anisotropic for the MRX experiments, in
which the magnetic fluctuations are observed.  This is because  the electron-ion 
collision rate is comparable to the frequencies and growth rates
of the mode, so it is just as accurate to take the pressure as isotropic.
Further, it is also appropriate to assume that the plasma is
isothermal, so that $ p= nT $ in general;  $ p_0 = n_0(x) T $
in the equilibrium and $ p_1 = n_1 T $ is the perturbation.
The fact that $ T $ is constant in the equilibrium over the region
occupied by the mode is supported directly from observations.
The fact that perturbations in the temperature are zero follows
from the very large thermal conductivity along the lines,
so that the thermal relaxation time is shorter than the perturbation
growth time. With this assumption we can avoid the solution
for $ f_0 $ and work entirely from the electron fluid equation,
to determine the perpendicular electron currents.

At this point we are in a position to solve for the ion and electron
currents in terms of the electric fields.  However, one further physical
result, charge neutrality, allows us to further shorten
the calculation.  Since the Debye length is very small compared to even
the electron gyration radius, we may assume to an excellent
approximation that the perturbed electron density $ n_{e1} $ is equal
to the perturbed ion density $ n_{i1} $ and this enable us to easily evaluate the
relevant terms in the perturbed equation of motions of the electrons.
(Of course if we had avoided this step and solved
directly for the ion and electron currents separately and then
substituted in Maxwell's equations, charge neutrality would have
followed automatically.  Introducing charge neutrality earlier leads
to considerable simplicity in the calculation and more physical
insight.)

	To summarize our calculation: we first write down the
equilibrium conditions.  Then next we calculate the perturbed
ion current and density from the unmagnetized ion dynamics.
We then calculate the perturbed perpendicular electron current from the
perpendicular equation of motion for the electrons.
We can then find the parallel electron current from
$ \nabla \cdot  {\bf  j}  = \nabla \cdot ({\bf  j}^i
+{\bf  j}^e) = 0 $.  Knowing these currents,
we then substitute them into Maxwell's equations
to find three independent relations for the wave electric fields.
However, it turns out that one of the three Maxwell's equations 
can be simplified to the electron force balance along the field line.
Thus, this eliminates the needs to calculate the parallel electron current
directly from $ \nabla \cdot {\bf  j} = 0$, which is demanded by
the charge neutrality condition.

\subsection{Equilibrium}

	For definiteness, we assume the MRX equilibrium
is a Harris equilibrium and study it in the ion frame.  
This seems the most physical frame in which
to study the instability since it turns out to be essentially
an unstable sound mode which is carreid by the ions.
We concentrate our attention on a small region say about half
way out from the center of the Harris sheet.

In this frame as shown in Fig.1(a), there is an electric field $ E_0 $
balancing their pressure force,
$T_i \partial n_0/\partial y$, in the $y$ direction:
\begin{equation}
en_0E_0=T_i {\partial n_0 \over \partial y}.
\label{i_balance}
\end{equation}
The magnetic field, $B_0$, is chosen in the $z$ direction.
A current is carried by electrons drifting in the $x$
direction with a speed $V_0$.
Force balance of the electron fluid then is given by
\begin{equation}
-en_0(E_0-V_0B_0)=T_e {\partial n_0 \over \partial y}
\label{e_balance}
\end{equation}
Eliminating ${\partial n_0 / \partial y}$ in
Eqs.(\ref{i_balance},\ref{e_balance}), we have
\begin{equation}
E_0={T_i \over T_e+T_i} V_0B_0.
\label{e0}
\end{equation}
If the plasma resistivity is finite, the electron current in the $x$ direction cannot be maintained
without an electric field in the same direction, $E_{x0}$. We shall see later, however, that
its effects on the wave dynamics are small as in the MRX.

\section{Dispersion Relation}

All wave quantities are assumed to have a normal mode decomposition proportional to
$$
\exp[i({\bf k} \cdot {\bf x} -\omega t)]
$$
with the wave vector ${\bf k}=(k_x,0,k_z)$ and the wave angular frequency $\omega$.
Note that ${\bf k}$ here does not have a $y$ component. This assumption
is justified in a local theory if wavelengths are much smaller than the
current layer thickness in the $y$ direction.

The governing equation between $\omega$ and ${\bf k}$, or the dispersion relation,
follows from three independent equations that relate the three components
of the wave electric field, $E_x$, $E_y$, and $E_z$. These can be derived
from Ampere's law and Faraday's law,
\begin{equation}
{\bf k} \times ({\bf k} \times {\bf E}) = -i\omega \mu_0 {\bf j},
\label{AmpFaraday}
\end{equation}
which leads to
\begin{eqnarray}
k_z^2 E_x - k_x k_z E_z & = & i\omega \mu_0 j_x
\label{eq1}\\
k^2 E_y & = & i \omega \mu_0 j_y
\label{eq2}\\
k_x^2 E_z - k_x k_z E_x & = & i\omega \mu_0 j_z.
\label{eq3}
\end{eqnarray}
Here $\mu_0$ is the vacuum magnetic permeability.
Next, we separately consider ion and electron dynamics to express the above 
equations in terms of the electric field.

\subsection{Ion Dynamics}

	We take the ions as unmagnetized and solve the kinetic
equation for the perturbed distribution function assuming the
equilibrium ion distribution function  is Maxwellian with constant
temperature, but variable density, $ d n_0/d y = \epsilon n_0 $.
From Eq.(\ref{i_balance}), $ \epsilon = e E_0/T_i = 2 e E_0/M v_i^2 $. 

The solution of the
ion Vlasov equation is carried out as an expansion to first order in
$ \epsilon $.
The result is most easily expressed in terms of the
electric field components $ E_1 $ and $ E_3 $ defined in Fig.1(b), in
which $ E_1 $ is the component parallel to $ {\bf k} $, and $ E_3 $
is the component perpendicular to it and in the $ x-z $ plane.
The perturbed ion current can then be written (Appendix B1),
\begin{eqnarray}
{\bf  j}^i = -i \frac{ n_0 e^2}{M} \frac{1}{k v_i}
 \left[ Z(\zeta) {\bf E} \right. & - &  ( \zeta  Z' + Z) ({\bf E} \cdot \hat{{\bf k} })
\hat{{\bf k} } \nonumber \\
& - &\left.  i (\epsilon/k) (\zeta  Z' + Z ) E_y \hat{{\bf k} }
\right]
\label{jip}
\end{eqnarray}
and the perturbed ion density is
\begin{equation}
n = i\frac{n_0 e}{M k^2 v_i^2} Z'(\zeta) \left( {\bf k} \cdot {\bf E}
+ i \epsilon E_y \right)
\label{ni}
\end{equation}
where  $ \hat{{\bf k} } ={\bf k}/k $, $ \zeta  = \omega /k v_i $, and
$Z$ is the plasma dispersion function.
We find that for the principal instabilities the phase velocity is
somewhat larger than $ v_i $ so for convenience we first take the
$ \zeta  \gg 1 $ limit (the cold limit), determine the parameter range of
instability. Then, in Appendix B3, we are able to employ a simple modification of the
dispersion relation to extract the correct growth rate including
the finite ion thermal effects.

In the cold limit the ion current neglecting the $ \epsilon $ correction is
obtained from the $ \zeta  \gg 1 $ limit and is
\begin{equation}
{\bf j}^i \approx i {\omega_{pi}^2 \over  \omega} \epsilon_0 {\bf E},
\label{ji}
\end{equation}
where the ion plasma angular frequency $\omega_{pi} \equiv
\sqrt{n_0e^2/M\epsilon_0}$ and $\epsilon_0$ is the vacuum susceptibility.
In the same limit the perturbed ion density is
\begin{equation}
n = i\frac{n_0 e}{M\omega^2}( {\bf k} \cdot {\bf E} +i \epsilon E_y )
\approx i{en_0 \over M\omega^2} ({\bf k} \cdot {\bf E}).
\label{n}
\end{equation}
The neglected $\epsilon$ term is much
smaller than the other one
since, for our local theory, we assume $k/\epsilon \gg 1$.
Indeed, it is shown in Appendix B2 that
the neglected term only has a small effect on the dispersion relation.

\subsection{Electron Dynamics}

As we have shown in Appendix A1, the perpendicular electron
current can be obtained
from the first order force balance for the electron fluid,
\begin{equation}
{\bf j}^e \times {\bf B}_0 = en_0 {\bf V}_0 \times {\bf B} + en_0 {\bf E} + en {\bf E}_0 + T_e\nabla n 
+ m n_0 {\partial {\bf  U}_E \over \partial t}
\label{je}
\end{equation}
where $ {\bf  U}_E = {\bf E} \times {\bf B_0} /B_0^2 $ and $m$ is the electron mass.
As shown in Appendix A2, the electron inertial terms contribute
a small effect to the distpersion relation and we can neglect
them when determining the instability.
The $y$ and $x$ components of Eq.(\ref{je}), therefore, are given by
\begin{eqnarray}
-j_x^e B_0 & = & -en_0V_0B_z + en_0E_y + enE_0 
 \label{jex} \\
j_y^eB_0 & =& en_0 E_x + ik_xT_en 
\label{jey}
\end{eqnarray}
respectively. Here, $B_z = k_x E_y/\omega$.
Since $E_0$ and $n$
are given already by Eq.(\ref{e0}) and Eq.(\ref{n}), $j_x^e$ and
$j_y^e$ can be expressed in terms of the electric field.

We note on the righthand side of Eq.(\ref{jey}) that there would be another term,
$enE_{x0}$, where $E_{x0}$ is the unperturbed electric field.
However, we will treat it as second order and balanced by quasi-linear terms~\citep{kulsrud04}.
In fact, the contribution from this term
is small when compared with the last term if $k_x \gg eE_{x0}/T_e$
as is often satisfied in the MRX.

The $z$-component of the electron current, $j_z^e$, is 
not determined by Eq.(\ref{je}).
It turns out, however, that it is unnecessary to explicitly calculate
it in order to obtain the dispersion relation due to simplifications of 
the $z$-component of Maxwell's equation, Eq.(\ref{eq3}).
This is because the electrons are so easily accelerated along the field line
by the force, $F_z^e$, on the electron fluid where
\[
F_z^e = -n_0 e \left( E_z + V_0 B_x + i k_z {T_e \over e} {n \over n_0}\right).
\]
The various terms in this force are separately large and must balance closely
to avoid very large parallel electrons currents.
In fact, taking $j_z^e= -n_0 e v_z^e = - i(e/m\omega) F_z^e$ 
and using Eq.(\ref{ji}) for the ion current, we can write
the $z$-component of Maxwell's equation, Eq.(\ref{eq3}), as
\begin{eqnarray}
&& k_x^2 E_z - k_x k_z E_x  =  i \omega \mu_0 (j_z^i + j_z^e) \nonumber \\
&& =  -{\omega_{pi}^2\over c^2} E_z  - {\omega_{pe}^2\over c^2} 
\left( E_z + V_0 B_x + i k_z {T_e \over e} {n \over n_0}\right). 
\label{eq3p}
\end{eqnarray}
Since $\omega_{pi}^2/\omega_{pe}^2=m/M \ll 1 $ and $(k\lambda_e)^2 \ll 1$
to our interests here, the above equation simplifies to the one
demanding the electron force balance in the $z$ direction,
\begin{equation}
E_z + V_0 B_y  + ik_z {T_e \over e} {n \over n_0} = 0,
\label{eq3a}
\end{equation}
where $B_y$ can be expressed in terms of the electric field using Faraday's law,
$$
B_y = {k_z E_x - k_x E_z \over \omega}.
$$
In Appendix A2, we show that the neglected terms have only a small effect on
the dispersion relation. We note that, although unneeded for the
dispersion relation, the $z$-component of the electron current, $j_z^e$,
can be determined by $\nabla \cdot {\bf j} =0$. This is a consequence of
the charge-neutrality condition, which is in turn enforced by 
Eq.(\ref{eq3a}).

It is interesting to note that if we allow the propagation angle approach to
$90^\circ$, the parallel phase velocity can be comparable to the electron thermal
velocity. In this case, we need to include a Landau term in Eq.(\ref{eq3a}).
Then, if $\beta_e \ll 1$, we would be able to recover the electrostatic perpendicular
LHDI~\citep{krall71}. 
However, since this electrostatic LHDI disappears at the high-$\beta$ of interest to us,
we need not include the Landau term.

\subsection{Dispersion Relation}

Substituting expressions of ${\bf j}^e$, ${\bf j}^i$, and $n$ [Eqs.(\ref{ji},\ref{jex},\ref{jey},\ref{n})]
into Eqs.(\ref{eq1},\ref{eq2},\ref{eq3a}) we obtain, after some algebra, the dispersion relation
\begin{equation}
\left(\begin{array}{ccc}D_{xx} & D_{xy} & D_{xz} \\D_{yx} & D_{yy} & D_{yz} \\D_{zx} & D_{yz} & D_{zz} \end{array}\right)
\left(
\begin{array}{c}
E_x \\
E_y \\
E_z
\end{array}
\right)=0,
\label{disp}
\end{equation}
where
\begin{eqnarray}
D_{xx}=& K^2\cos^2 \theta +1 -\displaystyle{\beta_i \over \beta_e +\beta_i} {KV \sin \theta \over \Omega}
\nonumber \\
D_{xy}=& i(\Omega-KV\sin\theta) \nonumber \\
D_{xz}=& -K^2\sin\theta\cos\theta-\displaystyle{\beta_i \over \beta_e +\beta_i} \displaystyle{KV \cos\theta\over \Omega} \nonumber \\
D_{yx}=&
-i\left(\Omega-\displaystyle{\beta_e\over 2}\displaystyle{K^2\sin^2\theta\over\Omega}\right) \nonumber \\
D_{yy}=& K^2+1 \nonumber\\
D_{yz}=& i\displaystyle{\beta_e\over 2}\displaystyle{K^2\sin\theta\cos\theta\over\Omega} \nonumber\\
D_{zx}=& KV\cos\theta -\displaystyle{\beta_e\over 2}\displaystyle{K^2\sin\theta\cos\theta\over\Omega}
\nonumber\\
D_{zy}=& 0 \nonumber\\
D_{zz}=& \Omega-KV\sin\theta -\displaystyle{\beta_e\over 2}
\displaystyle{K^2\cos^2\theta\over\Omega}. \nonumber\end{eqnarray}
Here the dimensionless parameters are defined by
\begin{aguleftmath}
\Omega \equiv {\omega \over \omega_{ci}}, \hspace{3mm}
K \equiv k{c\over \omega_{pi}}, \hspace{3mm}
V \equiv {V_0 \over V_A}, \hspace{3mm}
\beta_e \equiv {n_0 T_e \over B_0^2/2\mu_0}, \nonumber\\
\beta_i \equiv {n_0 T_i \over B_0^2/2\mu_0}, \hspace{3mm}
\sin \theta \equiv {k_x \over k}.
\label{parameter}
\end{aguleftmath}
Here, $\omega_{ci}$ is the ion cyclotron angular frequency
$eB_0/M$ and $V_A$ the Alf\'{v}en speed $B_0/\sqrt{\mu_0Mn_0}$.

The $ KV $ term in $ D_{{x x}} $ and in $ D_{xz} $ and the
$ \beta_e $ terms all result from replacing the kinetic equation for
the perturbed density $ n $ by it cold limit.  The 'one's in
$ D_{x x} $ and $ D_{yy} $ are ion currents which are similarly appoximated.

The resultant dispersion relation $\Omega(K)$ is a fourth order algebraic equation in $\Omega$
with 4 controlling parameters, $V$, $\beta_e$, $\beta_i$, and $\theta$,
\begin{aguleftmath}
\Omega^4 -2KV\sin\theta \Omega^3 \nonumber \\
-\left[(K^2+1)(K^2\cos^2\theta+1)
-K^2V^2\sin^2\theta+{\beta_e\over 2}K^2\right]\Omega^2
\nonumber \\
+KV\sin\theta \left[\beta_eK^2+(K^2+1){\beta_e+2\beta_i \over \beta_e+\beta_i}\right] \Omega
\nonumber\\
+K^2\left[ {\beta_e\over 2}\left[ (K^2+1)^2\cos^2\theta -K^2V^2\sin^2\theta\right] \right.\nonumber \\
\left. -(K^2+1)V^2{\beta_i \over \beta_e+\beta_i}\right]=0.
\label{dispersion}
\end{aguleftmath}

\section{Wave Characteristics and Instability}

\subsection{Basic Wave Characteristics without Drift}

The basic wave characteristics described by Eq.(\ref{dispersion}) are summarized here
for the case that there is no drift between ions and electrons. When $V=0$ and $\theta=0$,
Eq.(\ref{dispersion}) reduces to
$$
\left[\Omega^2-(K^2+1)^2\right]\left[\Omega^2-{\beta_e\over 2}K^2\right]=0
$$
which represents four waves, as shown in
Fig.2 for the case of $\beta_e=\beta_i=1$.
Two waves are whistler waves, traditionally termed fast waves, while
the other two waves are sound waves or slow waves. One of each waves propagates
along the background magnetic field and the other propagates against. As expected,
the whistler waves are largely transverse waves or electromagnetic waves since
the electric field vectors are perpendicular to the propagation (${\bf k}$) direction
$\phi \simeq 90^\circ$, where $\cos\phi \equiv {\bf k} \cdot {\bf E}/ (|{\bf k}||{\bf E}|)$.
In contrast, the sound waves are largely longitudinal waves or electrostatic waves
since $\phi \simeq 0$.

The situation changes when $\theta$ and $\beta$ are varied. In Fig.3,
the angles between ${\bf E}$ and ${\bf k}$, $\phi$, are shown for $V=0$ and a few cases of $\theta$
and $\beta$. It can be seen that when $\theta$ is larger, the whistler waves become
less electromagnetic and more electrostatic while the sound waves become
more electromagnetic and less electrostatic. This trend is stronger for larger values of
$\beta$.

\subsection{An Oblique Electromagnetic Instability}

It is evident that the whistler waves are supported by fast electron dynamics while
the sound waves are supported by slow ion dynamics. When there is no drift between
these two fluids, all wave branches stay separate in the dispersion diagram as shown
in Fig.2 for $\theta=0$. The situation is similar for more general cases 
of $\theta \neq 0$. If $V=0$, Eq.(\ref{dispersion}) reduces to
\begin{aguleftmath}
\Omega^4 - \left[(K^2+1)(K^2\cos^2\theta+1)+{\beta_e\over 2}K^2\right] \Omega^2 \nonumber\\
+{\beta_e\over 2}K^2(K^2+1)^2\cos^2\theta=0,
\label{dispnov}
\end{aguleftmath}
which represents four waves in the left panels in Fig.4 for the case of $\theta=60^\circ$ and $\beta_e=\beta_i=1$.  It is seen that at this propagation angle,
$\phi$ is $\sim 40^\circ$ for whistler waves and $\sim 0^\circ$ for sound waves. 

When there is a finite electron drift in the ion rest frame, 
the whistler waves are doppler-shifted so that each $\Omega$ from Eq.(\ref{dispnov})
is increased by $KV\sin \theta$, shown as dotted curves in the top-right panel of Fig.4 for the
case of $V=6$.
In constrast, sound waves, unaffected by the drift, are shown as dotted straight lines.
When the drift is large, some part of the backward propagating whistler waves branch can 
intercept with the forward propagating sound wave branch, resulting
in instabilities through reactive couplings. 
The case of $V=6$ is shown in the right panels of Fig.4 and all other parameters 
are the same as in the left panels.
It is seen that when $K < \sim 6$ or $K > \sim 16$, all four roots are real and thus all waves are
stable. When $6 < K <16$, two of roots become complex conjugates as
a result of coupling; one of them is damped and another growing 
(the growth rates are shown in the middle-right panel).
The maximum growth rate is about 8 times of $\omega_{ci}$ at $K\simeq 11$.
Since the polarization angle $\phi \simeq 15^\circ$, the unstable waves have significant
electromagnetic components.

Figure 5 shows the unstable region and contours of polarization angle in the
$\theta-K$ plane for a few values of $V$. 
It is seen that the unstable waves
are localized to small $K$ when $\theta$ is small and to large $K$ when $\theta$ is
large. The unstable region expands and the growth rate increases with increasing $V$.
The polarization angle $\phi$ ranges between $10^\circ$ to $25^\circ$,
and is larger near the small $K$ and small $\theta$ corner.

\section{A Physical Picture}

\subsection{Further Simplification of Electron Dynamics}

In order to understand the primary feedback mechanism
of our  instability we make further simplifications to the
dispersion relation given by Eq.(\ref{disp}).
We first start by rotating the coordinate for ${\bf E}$ as shown in
Fig.1(b): $(E_x,E_y,E_z)$ to $(E_1,E_2,E_3)$. $E_1$ is in the $k$ direction,
representing the electrostatic component. $E_2$ is the same as $E_y$ and $E_3$ is
another perpendicular component to ${\bf k}$, and both of these are
electromagnetic components. Using the new bases, $(E_1,E_2,E_3)$, Eq.(\ref{disp})
reduces to
\begin{equation}
\left(\begin{array}{ccc}D_{11} & D_{12} & D_{13} \\D_{21} & D_{22} & D_{23} \\D_{31} & D_{32} & D_{33} \end{array}\right)
\left(
\begin{array}{c}
E_1 \\
E_2 \\
E_3
\end{array}
\right)=0,
\label{dispprime}
\end{equation}
where
\begin{eqnarray}
D_{11}=& \sin\theta-\displaystyle{\beta_i \over \beta_e+\beta_i} {KV \over \Omega}
\nonumber \\
D_{12}=& i(\Omega-KV\sin\theta) \nonumber \\
D_{13}=& -(K^2+1)\cos\theta \nonumber \\
D_{21}=&
-i \displaystyle{\sin\theta \over \Omega} \left(\Omega^2-\displaystyle{\beta_e\over 2}K^2\right) \nonumber \\
D_{22}=& K^2+1 \nonumber\\
D_{23}=& i\Omega\cos\theta \nonumber\\
D_{31}=& \displaystyle{\cos\theta \over \Omega} \left(\Omega^2-\displaystyle{\beta_e\over 2}K^2\right) \nonumber\\
D_{32}=& 0 \nonumber\\
D_{33}=& \Omega\sin\theta-KV. \nonumber\end{eqnarray}

Again the $ KV $ term in $ D_{11} $ and the $ \beta_e $ terms
result form approximating the perturbed density, and the 'one's
in $ D_{13} $ and  $ D_{22} $ from approximating the ion currents.

Next we simplify these equations by taking the limit of large
$\Omega$, $K$, and $V$ since this asymptotic limit will make the
physical mechanism of the instability clear.
The simplified matrix then reduces to
\begin{equation}
\left(\begin{array}{ccc}
- \displaystyle{{\beta_i \over \beta_e+\beta_i} {KV \over \Omega}} & -iKV\sin\theta & -K^2\cos\theta\\
-i\displaystyle{\sin\theta \over \Omega} \left(\Omega^2- \displaystyle{\beta_e \over 2}K^2\right) & K^2 &  0\\
\displaystyle{\cos\theta \over \Omega} \left(\Omega^2- \displaystyle{\beta_e \over 2}K^2\right) & 0 & -KV\end{array}\right).
\label{dispred}
\end{equation}

Each line of the above matrix equation represents the balance of the leading forces
on the electron fluid along the
three coordinate directions $ y, x, z, $ respectively.
By referring back to Eq.(\ref{eq3}) and Eq.(\ref{je}) we can see that
the force balance can be written
\begin{equation}
\begin{array}{ccc}
y: & -enE_0 - j_{0x} B_z - j_x B_0 = 0\\
x: & -en_0E_1\sin\theta -\partial p_e/\partial x+ j_y B_0 =  0\\
z: & -en_0E_1\cos\theta -\partial p_e/\partial z + j_{0x} B_y = 0
\end{array},
\label{forces}
\end{equation}
where in the asymptotic limit the current ${\bf j}$ is all due to the electrons.
Interestingly, the electrostatic force is balanced by the Lorentz force in all directions.
In the $y$-direction, the unperturbed electrostatic field acting
on the perturbed electron density is balanced by the Lorentz force,
which consists of both magnetic pressure gradient,
$-j_{0x}B_z $, and tension $-j_x B_0$ forces. By contrast,
the perturbed electrostatic field is balanced by the magnetic tension,
$j_{0x}B_y$, in the $z$-direction.

\subsection{The Case of $\theta$=0}

We start with the simplest case, $\theta=0$, in which there are no perturbed forces 
in the $x$-direction. In the $y$-direction the perturbed magnetic pressure force is also zero
since $B_z=k_xE_y/\omega=0$. Therefore,
the electrostatic force, $-enE_0$, must be balanced by the magnetic tension force,
$-j_xB_0$. Suppose that the electron density is perturbed in a way such that $n>0$ at the origin
as illustrated in Fig.6(a) in the $y-z$ plane.
Because ${\bf E}_0$ points in the positive $y$-direction, the
perturbed electrostatic force on the electron fluid, $-enE_0$, 
points in the negative $y$-direction at the origin. Since it varies in $z$,
this force bends the field line until its 
magnetic tension force $-j_xB_0$ balances the
$-enE_0$ force. (Here the field-line bending can also be understood as
a result of the perturbed $j_x$ due to changing the number of the charged carriers
by the perturbed density $n$.)

In the $z$-direction, there is now a component of the magnetic tension force towards the origin  
$ j_{0x} B_y $ due to the bent line, as illustrated in Fig.6(a).
This force reduces or reverses the perturbed electrostatic force $ - e n_0 E_1$ 
produced by the electron density perturbation. In the {\it latter} case,
the perturbed electrostatic force is directed away from the regions
where $n>0$ and towards the regions where $n<0$.
As a result the perturbed electric field, ${\bf E}_1$, must point 
from the regions where $n<0$ to the regions where $n>0$, such as the origin.

To see that this leads to instability consider the ions
which only see the electrostatic field $ {\bf E}_1$.
This electrostatic field will force the ions to condense further at the origin
increasing their density perturbation.  By charge 
neutrality this will increase the initially assumed
electron density perturbation and thus lead to instability.

\subsection{The Case of $\theta > 0$}

We find that it is convenient to take the limit of $\beta_e=0$
for the discussion of this more general and complicated case.
Here the feedback to initial perturbations through compression or decompression of the electron
fluid along the $z$-direction is unaffected except for a reduced efficiency.
However, there are perturbed forces in the $x$-direction.
As before, we suppose an electron density perturbation $n>0$ at the origin.
When the mode is unstable, the perturbed electrostatic force, which is parallel to ${\bf k}$, 
has an $x$-component, $-en_0E_1\sin\theta$,
pointing away from the regions where $n>0$ towards the regions where $n<0$ also as before.
This force on the electrons decompresses the magnetic field in the $n>0$ regions and compresses it
in the $n<0$ regions. This is illustrated in Fig.6(b) in the $x-z$ plane.
Because ${\bf k}$ makes a finite angle to ${\bf B}_0$, 
the magnetic field lines are distorted to have both a tension force and also a magnetic 
pressure force.
Therefore, $B_z$ must be negative (decompressed) at the origin where $n>0$ and thus,
the associated magnetic pressure force in the $y$-direction, $-j_{0x}B_z$, is directed towards
positive $y$-direction. As a result, this force counters the initial electrostatic force, $-enE_0$,
(which bends the field line) and thus, reduces the tendency towards instability.

Both these stabilizing and destabilizing forces are included in the dispersion
relation from Eq.(\ref{dispred}), in which we restore $\beta_e$ to obtain
\[
\Omega^2={\beta_e\over 2} K^2 + {\beta_i \over \beta_e + \beta_i} 
{K^2V^2 \over V^2\sin^2\theta-K^2\cos^2\theta}.
\]
Consider a given (large enough) $V$, it can be seen that instability occurs when $K$ exceeds some threshold values, and 
stability returns eventually in the limit of large $K$, consistent with Fig.4.
Thus, if $\rho_e$ is small enough, the growth rate reaches its peak at a wavelength
longer then $\rho_e$. 
However, it is clear from the above equation that, if $\beta_e=0$, the instability persists over
all $K$ above its critical value (at least until some finite electron gyroradius effects
become important.)
From this, we can see that our calculation is essentially based on a two-fluid model,
and it is not strictly a Hall MHD calculation since the ions are totally unmagnetized
and one cannot set $\beta_e=0$ without losing the above physical contents.
Our calculation is perhaps closer to a hybrid model~\citep[see][]{birn01} 
with kinetic ions and a massless electron fluid, but in three dimensions.
We emphasize here that
the background ion pressure gradient is essential for the instability in both $\theta=0$ and
$\theta > 0$ cases because of the important
role played by the associated equilibrium electric field, ${\bf E}_0$.

\section{Discussions and Conclusions}

In the MRX, it has been observed that the usual electrostatic LHDI, propagating perpendicularly
to the magnetic field, is active only in the low-$\beta$ edge of the reconnection region, but
not in the high-$\beta$ central region~\citep{carter02a}.
This is consistent with the theoretical prediction that the perpendicular LHDI is stable at the high-$\beta$~\citep{davidson77,carter02b}. 
On the other hand, it has been found that, in the high-$\beta$ central region, 
obliquely propagating, electromagnetic waves in a similar frequency range are active, 
and their amplitude positively correlate with the reconnection rate~\citep{ji04a}.
Motivated by these observations,
we have developed a simple two-fluid formalism to derive and analyze in detail 
an electromagnetic drift instability in the lower-hybrid frequency range.
We term this the oblique LHDI.

We show that the main features of the instability are consistent with 
fully electromagnetic kinetic calculations~\citep{lemons77,wu83,tsai84}. We find that,
contrary to the perpendicular LHDI, the oblique LHDI
persists in high-$\beta$ plasmas. Further, the growth rate peaks
at longer wavelength than electron gyroradius, justifying our assumption that
the electrons are magnetized. The resultant waves
have mixed polarization and significant electromagnetic components.
The instability is caused by reactive coupling between the backward
propagating whistler (fast) waves in the moving electron frame and the forward propagating 
sound (slow) waves in the ion frame, and occurs when the relative drifts are large. 
After further simplifications of the model, the primary positive feedback mechanism 
is identified as a reinforcement of initial electron density perturbations by compression 
of the electron fluid by an induced Lorentz force. Interestingly, the revealed mechanism of the
instability requires close interactions between the electrostatic and electromagnetic forces.
In contrast to most of previous theories on MTSI, 
our analysis also suggest that the self-consistent background-ion-pressure
gradient is essential for the instability.

A few comments on three-dimensional particle simulations are in order.
In addition to the dimensionless parameters of Eq.(\ref{parameter}), 
the mass ratio, $M/m$, is another important parameter.
To make the simulations feasible, often $M/m$ is limited to a few hundreds.
In contrast, our analysis based on the above simple local model is valid in the limit
of large $M/m$ since ions are treated as unmagnetized.
Small mass ratios used in simulations will limit available wavenumber window
for the instability due to the condition of $\lambda_i^{-1} \ll k \ll \lambda_e^{-1}$.
In addition, the limited grid size and resolution may not permit numerical
treatment of the large oblique wavenumber range where our instability resides.
Future numerical simulations with increasingly powerful computers may help to 
elucidate these effects more clearly especially with regard to nonlinear consequences in magnetic
reconnection. Simulations of non-Harris current sheets, as attempted in the linear analyses~\citep{yoon04,sitnov04}, may prove to be more physically meaningful since they
may represent the reality more accurately.

Many of the predicted features of unstable waves discussed in this paper are also qualitatively consistent with
the observed magnetic fluctuations in the MRX~\citep{ji04a}, including their existence in the
high-$\beta$ region, their frequency range, and their propagation direction with respect to
the background magnetic field. 
In fact, the parameters we use in the calculation have been drawn directly from
the MRX experiments, and they are valid throughout the bulk of the MRX current sheet. 
Also, the instability does indeed persist into the $\beta_e\gg 1$ regimes, but
the physics of the instability is still uncertain in the region where the magnetic field 
nearly vanishes.
One particular comment on their phase velocity is worth making.
The experimentally measured phase velocity is of the same order as the relative drift velocity.
Even given the large experimental uncertainties such as the measurement location and the unknown
relative velocity between the ion frame and the laboratory frame,
the measured phase velocities are considerably larger than our theoretical predictions.
As seen in Fig.4, the unstable waves should have phase velocities 
on the order of the ion thermal speed. However, the theory presented here is limited
to the case where $k_y=0$.
The phase velocity may be substantially increased by incorporating a nonzero $k_y$.
This is a subject for future work. 
Increasing the phase velocity to values much larger than ion thermal speed 
may also help mitigate another shortcoming of our analysis:
the reduction of the growth rates by ion thermal effect.
The role which this instability plays in magnetic reconnection, such as in the production of
anomalous resistivity and its effect on heating, is discussed in
a forthcoming paper~\citep{kulsrud04} that is based on quasi-linear theory.

\begin{appendix}

\section{Detailed Calculations of Electron Dynamics}

\subsection{Drift Kinetic Equation for Electrons}

Normally, the drift kinetic equation is developed for both
electrons and ions, and is combined with Maxwell's equations
to achieve some important simplifications.  This full
formulation is described in a number of places, for example
in \cite{kulsrud83}. However, if the
ions are unmagnetized, as in this paper, the formulation is reduced to
that of solving the electron Vlasov equation alone, as an expansion
in $ \rho_e/\lambda $, and $ 1/\omega_{ce} t $ where $ \lambda $ is
the length scale of the phenomena, and $ t $ is its time scale.
We follow the procedure given in the handbook article.  It is clear that
the electronic charge can be used as a guide to the expansion and we use
$ 1/e $ as the expansion parameter.

The electron Vlasov equation is
\begin{equation}\label{A1}
\frac{\partial f}{\partial t} +{\bf  v} \cdot \nabla f -
\frac{e}{m} \left( {\bf E} + {\bf  v} \times {\bf B} \right)
\cdot \nabla_{{\bf  v}} f = 0.
\end{equation}
We first carry out the expansion for the full distribution,
(equilibrium $ f $ and perturbed $ \delta f $) and later carry out the
expansion in the instability perturbation.

The lowest order Vlasov equation is accordingly
\begin{equation}\label{A2}
- \frac{e}{m} \left( {\bf E} + {\bf  v} \times {\bf B}  \right) \cdot
\nabla_{{\bf  v}} f_0 = 0.
\end{equation}
We introduce the $ {\bf E} \times {\bf B} $ velocity by
\begin{equation}
{\bf  {\bf U}_E} = \frac{{\bf E} \times {\bf B} }{B^2}
\end{equation}
and carry out the transformation of the velocity at each point
$ {\bf  r} $,
\begin{equation}
{\bf  v} ={\bf  {\bf U}_E}({\bf r}) +{\bf  v'} =
{\bf  U}_E +v_{\perp} \cos \phi \hat{{\bf  x'}}  +
v_{\perp} \sin \phi \hat{{\bf  y'}} +  v_{\parallel } {\bf  b}
\end{equation}
where $  \hat{{\bf  x'}}  , \hat{{\bf  y'}} $ and $ {\bf  b} $
are local coordinates at each point $ {\bf  r} $, and $ v_{\perp },
 \phi $ and $ v_{\parallel} $ are cylindrical coordinates for
$ {\bf  v'} $.  Then Eq.(\ref{A2}) becomes
\begin{equation}
 \frac{e B}{m} \frac{\partial f_0}{\partial \phi } -
\frac{e E_{\parallel }}{m}  \frac{\partial f_0}{\partial v_{\parallel}}
=0.
\end{equation}
If $ E_{\parallel } $ is non zero, $ f_0 $ would be constant along
a helical orbit in velocity space that extends to infinity, which is
impossible.  Thus, $ E_{\parallel} $ must vanish to lowest order and
$  E_{\parallel} $ must be considered first order.

Dropping the second term we see that $ f_0 $ is independent of
$ \phi $ (gyrotropic) and, thus, a function only of $ t, {\bf  r},
v_{\perp} , $ and $  v_{\parallel} $.

Proceeding to next order in $ 1/e $ we get
\begin{equation}\label{A6}
-\frac{eB}{m} \frac{\partial f_1}{\partial \phi} =
\left(\frac{\partial f_0}{\partial t} + {\bf  v} \cdot \nabla f_0 \right)
- \frac{e}{m} E^1_{\parallel} \frac{\partial f_0}{\partial v_{\parallel} }
\end{equation}
where the expression in parentheses must be transformed to
$ t, {\bf  r} ,v_{\perp}, v_{\parallel}, \phi $ coordinates.

	Equation (\ref{A6}) can only be solved for $ f_1 $ if its average over
$ \phi $ (which eliminates $ \partial f_1/\partial \phi $ ) vanishes.
The result is

\begin{eqnarray}
\frac{\partial f_0}{\partial t} & + &
( {\bf  U}_E  +   v_{\parallel} {\bf  b} )
\cdot \nabla f_0 \nonumber \\
& - & \frac{v_{\perp} }{2} \left( \nabla \cdot {\bf  U}_E
- {\bf  b \cdot \nabla U}_E \cdot {\bf   b} +
v_{\parallel} \nabla \cdot {\bf  b} \right)
\frac{\partial f_0}{\partial v_{\perp} }  \nonumber\\
& + & \left(  - {\bf  b} \cdot \frac{D {\bf  U}_E}{D t} \cdot {\bf  b}
+\frac{v_{\perp}^2}{2} (\nabla \cdot {\bf  b } )
 + \frac{e}{m} E_{\parallel} \right) \frac{\partial f_0}{\partial v_{\perp} }
= 0 \nonumber \\
&& \label{A7}
\end{eqnarray}
where $D{\bf U}_E/Dt \equiv \partial {\bf U}_E /\partial t +
( {\bf U}_E+{\bf b}v_\parallel)\cdot \nabla  {\bf U}_E$.
(Note that the $ e E_\parallel $ term is zero order since $ E_{\parallel} $
is first order and $ e $ is minus first order.)

	In principle, $ f_0 $ can be solved for from this equation.
For the case of the instability, $ f_0 $ can be written as $ f^0_0 +
\delta f_0 $ where $ f_0 $ is a local Maxwellian, $ {\bf U}_E $ is
a perturbation, and $ {\bf  B}_0 = B_0 \hat{{\bf  z} } $ to lowest order.
 The only equilibrium term that survives is the $ v_{\parallel}
{\bf  b} \cdot \nabla f_0^0 $ term so the only restriction
on $ f^0_0 $ is that it be constant along the magnetic field.

	To get the electron current perpendicular to $ {\bf B} $ we need $ f_1 $,
\begin{equation}\label{A8}
{\bf  j}_{\perp}^e = e \int {\bf v}_{\perp} f_1  d \phi v_{\perp} d v_{\perp}
d v_{\parallel} = e \int \frac{\partial {\bf v}_{\perp} }{\partial \phi }
\frac{\partial f_1}{\partial \phi }  d^3 {\bf  v}
\end{equation}
which can be obtained directly from Eq.(\ref{A6}) by multiplying it
by $ \partial {\bf v}_{\perp} /\partial \phi =
- v_{\perp} \sin \phi  \hat{{\bf  x} }
 + v_{\perp} \cos  \phi  \hat{{\bf  y} } $, dividing by $ B $ and inegrating over
velocity space.  In fact, we could just as well have multiplied Eq.(\ref{A6})
by $ {\bf v}_{\perp} $ and integrated it to find $ {\bf  j}_e \times {\bf B} $.
Even simpler, we could have multiplied Eq.(\ref{A1}) by $ {\bf B} $ integrated
over velocity space and taken the perpendicular part of the result.
This result would be the perpendicular part of
\begin{equation}\label{A9}
n m \left( \frac{\partial {\bf  v}_e}{\partial t } +
{\bf  v}_e \cdot \nabla {\bf  v}_e \right) = {\bf  j}^e \times {\bf B}
- \nabla \cdot {\bf  P}^e + n e {\bf E}
\end{equation}
Here the stress tensor is zero order, and can be found from $ f_0 $
once we have solved Eq.(\ref{A7}) for it.

If we inspect Eq.(\ref{A9}) we see that the inertia term and the
$ \nabla \cdot {\bf  P} $ are zeroth order, but the $ n e E $ term is minus
first order in the $ 1/e $ expansion.  Thus, $ {\bf  j}^e $ has a minus
first order part, $ n e $ time the $ {\bf E} \times {\bf B} $ drift, and
 zero order parts, essentially the diamagnetic and polarization currents.
If the ions were magnetized, this minus first order current would be
cancelled by the corresponding $ {\bf E} \times {\bf B} $ current of the
ions, but this is no longer the case for unmagnetized ions.

This procedure gives the perpendicular current of the electrons.  The
parallel current is given by the continuity condition
\begin{equation}
\nabla \cdot {\bf  j}^e + \frac{\partial }{\partial t}
(n e) = 0
\end{equation}
Again for finite $ n $ the $ ne $ term is minus first order.  $ n_0 $
is given by the zero moment of $ f_0 $. However, $ n_1 $ is needed to give
the finite parallel electron current, and for it we need the zero
moment of $ f_1 $.  This zero moment cannot be obtained from Eq.(\ref{A6}), 
which only gives the $ \phi $ dependent part of $ f_1$, $\partial f_1/
\partial \phi $. To get the mean part it is necessary to go to next order in the
$ 1/e $ expansion of the Vlasov equation.  This has been done some time
ago~\citep{frieman66}, and will yield $ n_1 $.

This procedure is certainly possible to carry out in all detail as outlined
above and is fairly easy   for our  perturbation problem. In fact if it is carried
out in a velocity frame in which the equilibrium electric field is zero
(the so-called Harris frame) the results turn out to be essentially identical
to those calculated by \cite{yoon94} in the common limit of approximation,
small gyration  radius and small frequency compared to the electron cyclotron
frequency.  As stated in the text, we can avoid some of the calculation by
taking the perturbed density from that of the ions by quasi neutrality.
This also avoids going to next order in the Vlasov equation to find $ n_1 $.
This assumption puts a constraint on $ E_{\parallel} $ in an early phase in the calculation
rather than waiting for substitution in Maxwell's equations to enforce
it. In any event the drift kinetic approach is completely consistent
with earlier calculations of LHDI.

\subsection{Electron Inertial Terms}

The first two rows of the matrix in Eq.(\ref{disp}) represent 
$-\Omega/n_0e$ times the $y$ and $x$ components of Eq.(\ref{je}).
Their initial terms are $i(m/M) n_0e\Omega E_x$ and
$-i(m/M) n_0e\Omega E_y$, respectively.
Multiplying these by $(-\Omega/n_0e)$ we get for the first two
rows of the matrix equation
\[
\begin{array}{ccc}
D_{xx} - \displaystyle{\frac{m}{M}} \Omega^2  &D_{xy} &D_{xz} \\
D_{yx} & D_{yy} + \displaystyle{\frac{m}{M}} \Omega^2  & D_{yz}
\end{array}
\]
where the coefficients $D$ are given by Eq.(\ref{disp}) as before.
The last row represents $-\Omega c^2/\omega_{pe}^2$ times
the last term in Eq.(\ref{eq3p}). Bringing all the other terms
to the right-hand side and multiplying these by $-c^2/\omega_{pe}^2$,
we get $-(m/M)K^2\sin\theta\cos\theta E_x +(m/M) (1+K^2\sin^2\theta)E_z$.
Multiplying these by $\Omega$ and adding the results to the last row of the 
matrix equation, we obtain 
\[
D_{zx} -\displaystyle{\frac{m}{M}} K^2 \Omega \sin \theta \cos \theta  \quad
D_{zy} \quad
D_{zz} +\displaystyle{\frac{m}{M}}\Omega(1+K^2 \sin^2 \theta).
\]

Transforming to the $(E_1, E_2, E_3)$ components of the electric field, we have
\[
\begin{array}{ccc}
D_{11} - \displaystyle{\frac{m}{M}}\Omega^2\sin\theta  & D_{12} & D_{13} +i\displaystyle{\frac{m}{M}} \Omega^2 \cos \theta \\
D_{21}   & D_{22} + \displaystyle{\frac{m}{M}} \Omega^2 & D_{23}  \\
D_{31} + \displaystyle{\frac{m}{M}}\Omega \cos \theta  & D_{32}   & D_{33} + \displaystyle{\frac{m}{M}} \Omega \sin \theta (1+ K^2)
\end{array}
\]
and for the limit of large $ K , V $ and $ \Omega $, Eq.(\ref{dispred}) becomes
\[
\left(\begin{array}{ccc}
- \displaystyle{{\beta_i \over \beta_e+\beta_i} {KV \over \Omega}} - \displaystyle{\frac{m}{M}}\Omega^2\sin\theta
& -iKV\sin\theta & -K^2\cos\theta +i\displaystyle{\frac{m}{M}} \Omega^2 \cos \theta \\
-i\displaystyle{\sin\theta \over \Omega} \left(\Omega^2- \displaystyle{\beta_e \over 2}K^2\right) & 
K^2 + \displaystyle{\frac{m}{M}} \Omega^2 &  0\\
\displaystyle{\cos\theta \over \Omega} \left(\Omega^2- \displaystyle{\beta_e \over 2}K^2\right) 
+ \displaystyle{\frac{m}{M}}\Omega \cos \theta & 0 & -KV
+ \displaystyle{\frac{m}{M}} \Omega \sin \theta (1+ K^2)
\end{array}\right).
\]

If we regard $ K, V $, and $ \Omega $ as all of order $ K $, then we can
we can see that the relative corrections are of order at most $ m/M $
except in the one-one and three-three elements where they are of order
$\sim (m/M) K $.  These corrections are all small and
can be neglected as long as $ K \ll M/m $.

	Incidentally,  the correction in the third
line represents the extra parallel electron field needed to
accelerate the electrons along the magnetic field to achieve
charge neutrality.  Its smallness indicates the ease with which
the electrons are able to achieve charge neutrality.

\section{Detailed Calculations of Ion Dynamics}

\subsection{Perturbed Ion Current and Density}

	The expressions for the unmagnetized
ion current and density given in Eqs. (\ref{jip}) and (\ref{ni}), which keep the equilibrium density 
gradient, as a first order  correction
are found from the perturbed ion distribution function
with the same correction.  The latter is obtained
by iterating the perturbed ion Vlasov equation
\begin{equation}
-i(\omega - {\bf k} \cdot {\bf  v} ) f_1 
+ v_y \frac{\partial f_1}{\partial y} +
\frac{e}{M} E_0 \frac{\partial f_1}{\partial v_y} 
+ \frac{e}{M} {\bf E}_1 \cdot \frac{\partial f_0}{\partial {\bf  v} }  
=0
 \end{equation}

The second and third terms are the correction terms.  
Therefore,  drop them at first and solve for the 
uncorrected $ f_1 $ from the remaining equation,
  in the standard way.

\begin{equation} 
f_1 = \frac{-i n_0 e}{k M v_i^5} 
\frac{2 {\bf  v} \cdot {\bf E}_1 }{(v_z - \omega/k )} 
\frac{e^{-v^2/v_i^2}}{\pi^{3/2}} 
\end{equation}
where $ v_i^2= 2 T_i/M $, and where without loss of 
generality we take the $ z $ axis along $ {\bf k} $.

	We see that $ \partial f_1/\partial y =
\epsilon f_1 $ where $ \epsilon = (d n_0/d y)/n_0 $.

Next, we insert this expression into the second
and third terms of the full Vlasov equation 
and solve for the correction, $ \delta f $, to 
$ f_1 $ which satisfies
\begin{equation}
-i(\omega - {\bf k} \cdot {\bf  v} ) \delta f = 
- v_y \epsilon f_1 - \frac{e}{M} E_0 
\frac{\partial f_1}{\partial v_y} 
\end{equation} 

After some algebra we can express the zero and first moments of 
$ (f_1 + \delta f ) $ in terms of the plasma dispersion 
 function, $ Z $,  of $ \zeta = \omega /k v_i $ and,
thus,  obtain Eqs.(\ref{jip}) and (\ref{ni}) of the main text.  

\subsection{Dispersion Relation with the Correction
from Background Density Gradient}

In Eq.(\ref{n}), a term proportional to the density gradient has been neglected
in deriving the dispersion relation. It is straightforward to show that, by including this term,
the dispersion matrix is given by
\begin{equation}
\left(\begin{array}{ccc}
D_{xx} & D_{xy} + 2i \displaystyle{{\beta_i \over (\beta_e+\beta_i)^2} {V^2 \over \Omega} } & D_{xz}\\
D_{yx} & D_{yy} + \displaystyle{{\beta_e \over \beta_e+\beta_i} {KV\sin \theta \over \Omega}} & D_{yz}\\
D_{zx} & D_{zy}+ i\displaystyle{{\beta_e \over \beta_e+\beta_i} {KV\cos \theta \over \Omega}} & D_{xz}
\end{array}\right),
\label{dispgrad}
\end{equation}
where the coefficients $D$ are given by Eq.(\ref{disp}). The resultant dispersion relation
remains as a fourth order equation, and the added new terms only have a small effect
on the solutions.
In the right and middle panel of Fig.4, the growth rate by Eq.(\ref{dispgrad})
is shown as the dotted line, which differs little from the solid line by Eq.(\ref{disp})
especially in the large $K$ limit.
(The dotted line indicating instability at very small $K$ has no physical significance
since the local approximation becomes clearly questionable for such cases.)

\subsection{Growth Rates with Warm Ions}

The most important instabilities occur for very local perturbations
with large $ K,V $ and $ \Omega $.  We restrict the discussion
of the thermal corrections to this case.

	The cold ion approximation involves using Eq.(\ref{n})
for the ion density instead of Eq.(\ref{ni}) and Eq.(\ref{ji}) for the ion currents
instead of Eq.(\ref{jip}).  In equation (18) the ion currents are negligible and
only the one-one element and the $ \beta_e $ terms  are proportion to the
perturbed ion density.  Thus the matrix of Eq.(\ref{dispred}) with the corrected ion density
is
\begin{equation}
\left(\begin{array}{{ccc}}
-\alpha \displaystyle{\frac{\beta_i }{\beta_e + \beta_i}} KV   &-i KV \sin \theta
& - K^2 \cos \theta   \\
-i \sin \theta \left(\Omega^2 - \displaystyle{\frac{\beta_e}{2}} K^2 \alpha \right) &K^2 & 0 \\
\cos \theta \left(\Omega^2 - \displaystyle{\frac{\beta_e}{2}} K^2 \alpha \right) & 0 & -KV \\
\end{array}\right)
\label{B1}
\end{equation}
where
\begin{equation}
\alpha = \frac{n_{true}}{n} = \zeta ^2 Z'(\zeta )
\end{equation}
where $ \zeta  = \omega /(k v_i) = \Omega/ (K \sqrt{\beta_i}) $.

The dispersion relation from Eq.(\ref{B1}) can thus be written
\begin{eqnarray}
- K^4V^2 \frac{\beta_i}{\beta_e + \beta_i } \alpha &+&
 K^2V^2 \sin^2 \theta\left( \Omega^2 - \alpha \frac{\beta_e}{2} K^2\right) \nonumber\\
&-& K^4 \cos^2  \theta \left( \Omega^2 - \alpha \frac{\beta_e}{2} K^2\right) =0.
\end{eqnarray}
By dividing this equation by $ \alpha $ we see that $ \Omega^2/\alpha $
satisfies the same equation as $ \Omega_0^2 $, the approximate solution for
the growth rate with cold ions.
Thus we can write
\begin{equation}\label{correct}
\frac{\zeta ^2}{\zeta ^2 Z'(\zeta)  } = \frac{1}{Z'(\zeta)  } =
\zeta^2_0
\end{equation}
where $ \zeta_0 ^2 = \Omega^2_0/(K^2 \beta_i). $ Thus from Eq.(\ref{correct})
we plot ratio of the true $\zeta$ to the approximate $ \zeta _0 $ as a function of $\zeta_0$
in Fig.7.
(Actually, the $ \Omega $'s are pure imaginary so we plot
$\Gamma/(K \sqrt{\beta_i}) $'s where the $\Gamma $'s refer to
approximate and exact normalized growth rates.)

We see that there is indeed a difference of order unity
between the approximate and exact values of $ \Omega $ or $ \zeta  $.
Since we see from Fig.4 that the peak $\Gamma_0/K \approx 1$,
the true $ \Gamma/K \approx 0.5 $ when $ \beta_i =1$ and $\Gamma/K \approx
0.7$ when $\beta_i = 0.5 $.  In spite of this reduction, we see that the
oblique LHDI is still unstable.

\end{appendix}


\begin{acknowledgments}
The authors are grateful to Mr. Y. Ren for his
contributions to initial assessments of wave dispersion of high-$\beta$ plasmas.
Drs. P. Yoon, A. T. Y. Lui, W. Daughton, and M. Sitnov are acknowledged for 
useful discussions. This work was jointly supported by DOE, NASA, and NSF.
\end{acknowledgments}

%
%
%
%
%
%
%
%


\begin{figure*}
\centerline{\includegraphics[width=2.5truein]{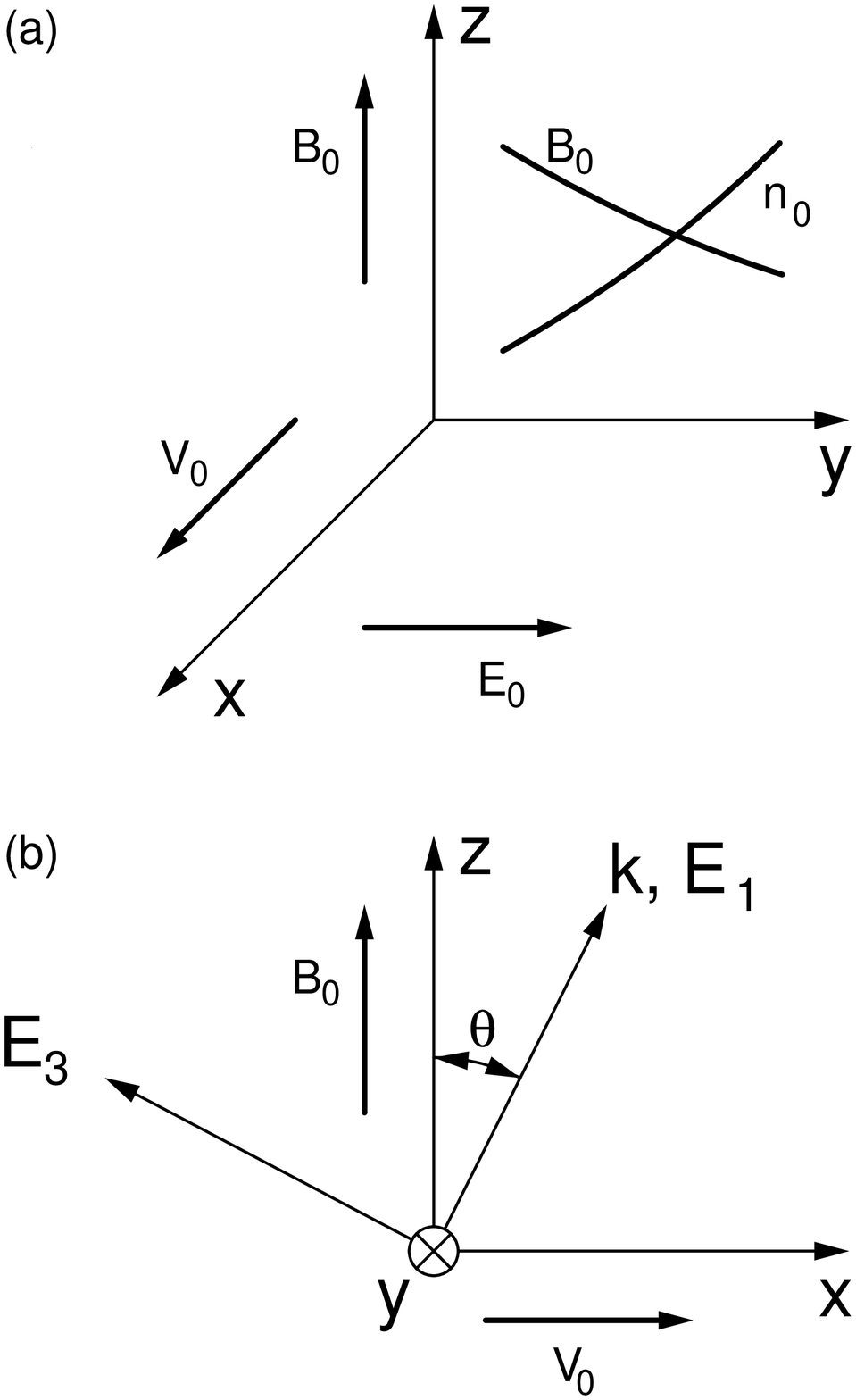}}
\caption{(a) Illustrations of the equilibrium state. Ions are at rest while electrons drift toward positive $x$ direction, crossing magnetic field in the $z$ direction. The resultant Lorentz force and electric field is balanced by pressure gradients in the $y$ direction, which points towards the current sheet center.
(b) Definitions of $E_1$ and $E_3$. $E_2$ is same as $E_y$.}
\label{equil}
\end{figure*}

\begin{figure*}
\centerline{
\includegraphics[width=3.0truein]{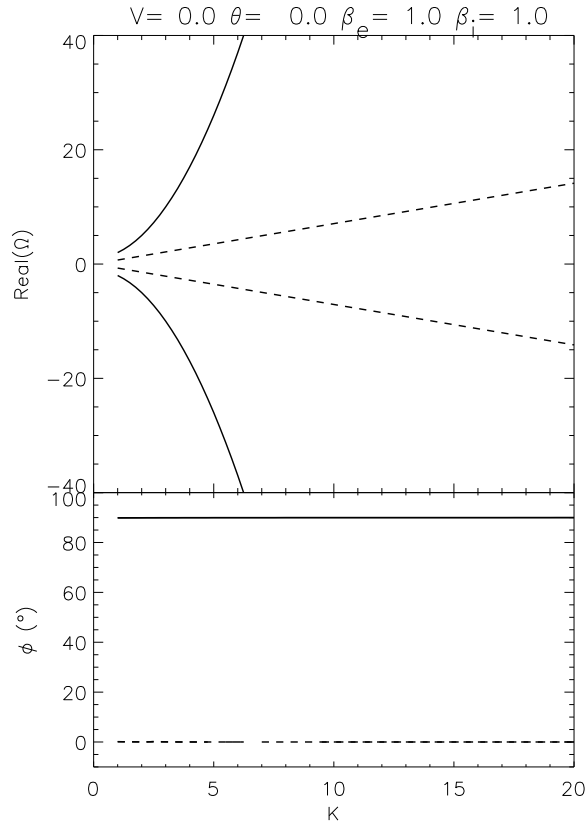}}
\caption{(top) Dispersion relation for the case that $V=\theta=0$ and
$\beta_e=\beta_i=1$. There are 2 whistler (fast) waves (solid lines) and
2 sound (slow) waves (dotted lines). (bottom) Angle ($\phi$) between ${\bf E}$
and ${\bf k}$ vector for both whistler waves (solid line) and sound
waves (dotted line).}
\label{4wave}
\end{figure*}

\begin{figure}
\centerline{
\includegraphics[width=5.0truein]{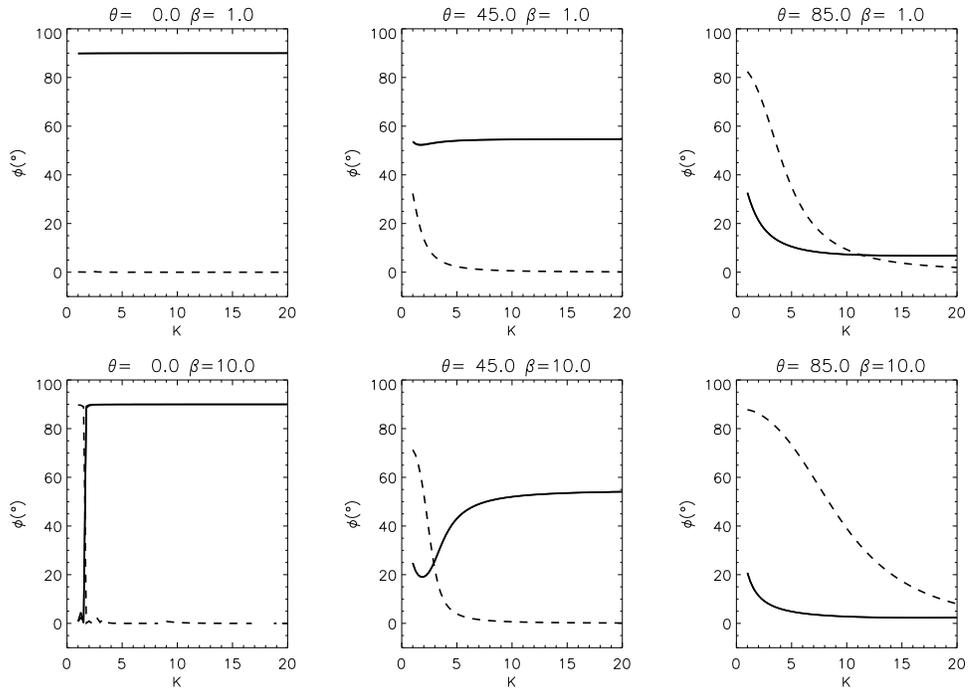}}
\caption{Angle between ${\bf E}$ and ${\bf k}$ for the cases of $\theta=0,45^\circ,85^\circ$
and $\beta_e(=\beta_i)=1$ and 10. Solid lines represent whistler waves and dotted lines represent
sound waves.}
\label{esem}
\end{figure}

\begin{figure*}
\centerline{
\includegraphics[width=5.5truein]{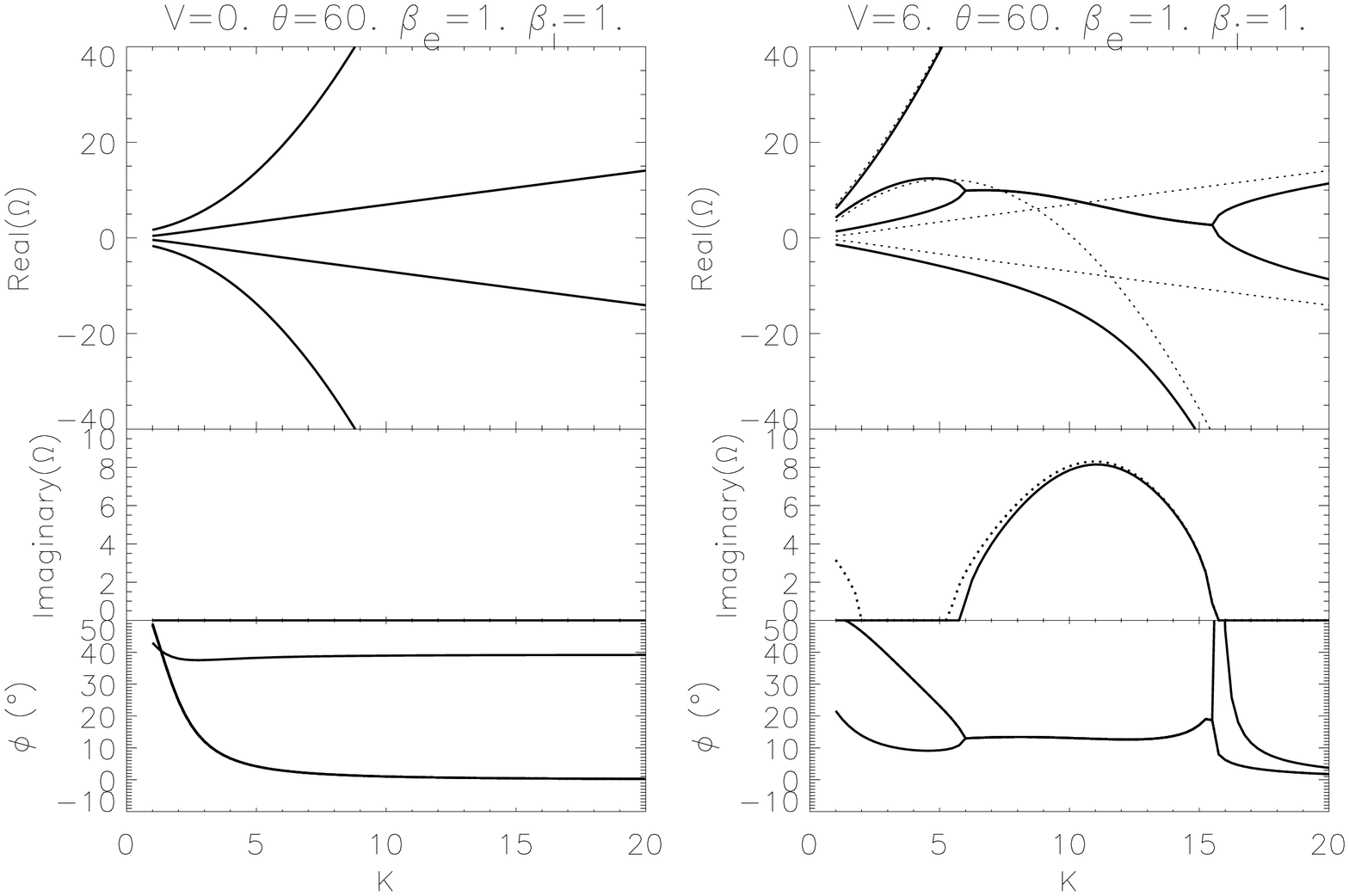}}
\caption{Dispersion relation (top) for the case of no drift (left) and large drift (right).
Growth rate (middle) and $\phi$ (bottom) are also shown for both cases. See main
text for the detailed explanation of the dotted lines.}
\label{inst}
\end{figure*}

\begin{figure}
\centerline{\includegraphics[width=6truein]{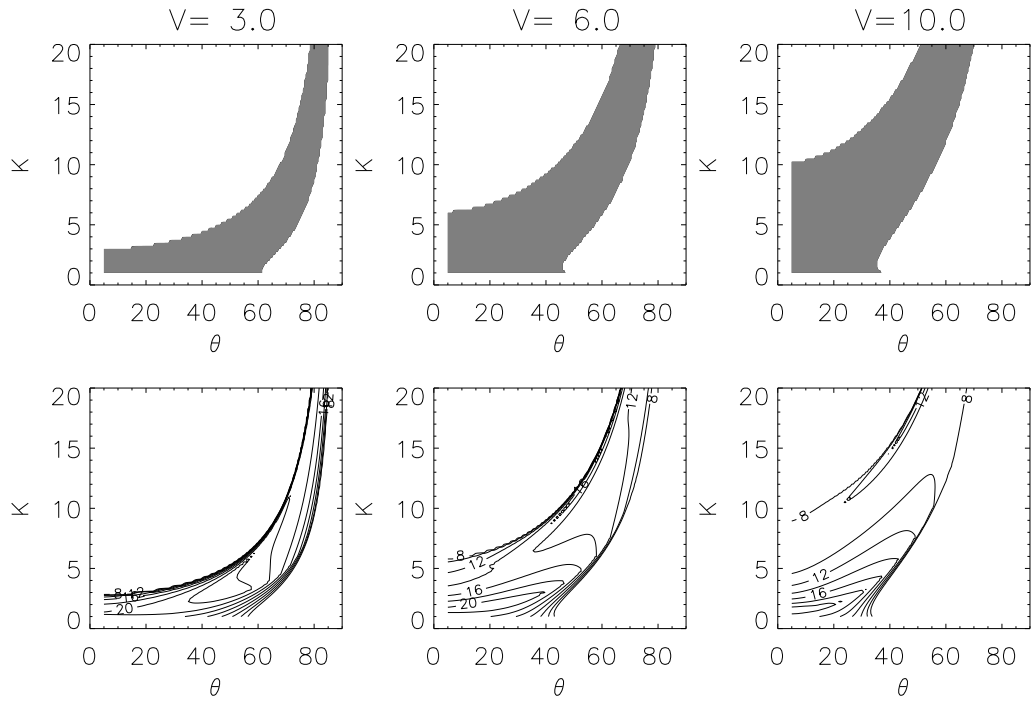}}
\caption{Unstable region where Im$(\Omega)>0$ (filled regions in top panels)
and contours of polarization angle ($\phi$, bottom panels)
in the $\theta-K$ plane for the cases of $V=3,6,10$ and $\beta_e=\beta_i=1$.}
\label{thetak}
\end{figure}

\begin{figure*}
\centerline{
\includegraphics[width=5.5truein]{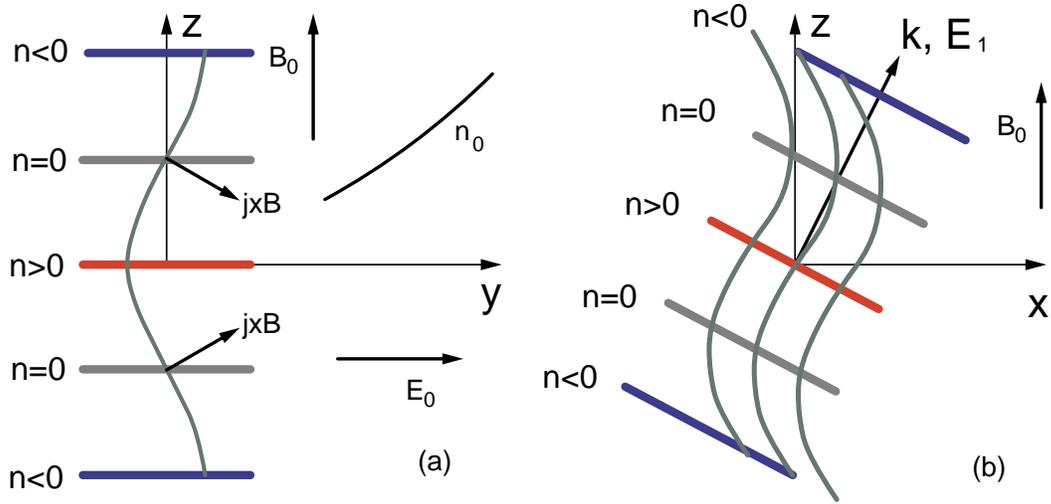}}
\caption{Illustrations of the instability mechanism (a) when $\theta=0$ in the $y-z$ plane
and (b) when $\theta>0$ in the $x-z$ plane. This distance between the field lines 
indicates the relative field strength.}
\label{picture}
\end{figure*}

\begin{figure*}
\centerline{\includegraphics[width=3.5truein]{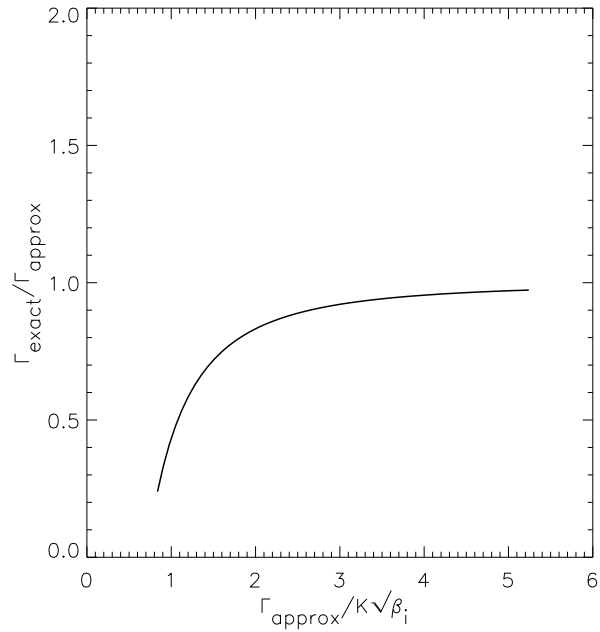}}
\caption{Ratio of exact growth rate to approximate growth rate
as a function of $\Omega/(K\sqrt{\beta_i})$.}
\label{ion-landau}
\end{figure*}

\end{article}

\end{document}